\documentclass[12pt]{article}
\usepackage{amstex}

\newfont{\zapfxii}{eusm10 scaled\magstep1}
\newcommand{\zap}[1]{\mbox{\zapfxii #1}}

\let\mb\mathbf
\let\bs\boldsymbol
\let\d\partial

\newcommand{\al}{\alpha}
\newcommand{\be}{\beta}
\newcommand{\ga}{\gamma}
\newcommand{\de}{\delta}
\newcommand{\De}{\Delta}
\newcommand{\ep}{\epsilon}
\newcommand{\ph}{\varphi}
\newcommand{\la}{\lambda}
\newcommand{\La}{\Lambda}
\newcommand{\PSI}{\bs\psi}
\newcommand{\n}{{\cal N}}
\newcommand{\p}[1]{{\cal P}_{#1}}
\newcommand{\pk}[1]{\zap P^{(k)}_{#1}}
\newcommand{\pp}[1]{\zap P^{(2)}_{#1}}
\newcommand{\RP}{\Bbb R{\rm P}^1}
\newcommand{\Dx}{\partial _x}
\newcommand{\Dz}{\partial _z}
\def\dd#1{\frac\partial{\partial #1}}
\newcommand{\drm}[1]{\zap{D}^1(#1)}
\def\g{\frak g}
\def\s{\frak s}
\def\hg{\hat{\frak g}}
\def\sl2{\frak{sl}(2)}
\def\s{\frak s}

\newcommand{\bpm}{\begin{pmatrix}}
\newcommand{\epm}{\end{pmatrix}}

\newcommand{\sch}{Schr\"o\-ding\-er}
\newcommand{\qes}{quasi-exactly solvable}

\newcommand{\fd}{finite-dimensional}
\newcommand{\dfo}{differential operator}
\newcommand{\QED}{\quad Q.E.D.}

\def\Je#1{J^\ep_{#1}}
\def\Jp#1{J^+_{#1}}
\def\J0#1{J^0_{#1}}
\def\Jm{J^-}
\newcommand{\Tk}{T^{(k)}}
\def\Ah{\hat A}
\def\Ut{\tilde U}
\def\Wt{\tilde W}
\def\Te{T^\ep}
\def\Tp{T^+}
\def\T0{T^0}
\def\Tm{T^-}
\def\Qa{Q_\al}
\def\M{M_{\al\be}}

\def\Tb{\,\overline{\!T}{}}
\def\Xb{\,\overline{\!X}{}}
\def\nb{\,\overline{\!\n}{}}
\def\zb{\bar z}
\def\gb{\bar\g}
\def\pb{\bar p}
\def\Qb{\,\overline{\!Q}{}}
\def\Qab{\,\overline{\!Q}{}_{\al}}
\def\qb#1#2#3{\,\overline{\!q}{}_{#1}(#2,#3)}
\def\qab#1#2{\qb{\al}{#1}{#2}}

\newcommand{\diag}{\operatorname{diag}}
\newcommand{\Span}{\operatorname{Span}}
\newcommand{\GL}{\operatorname{GL}}
\newcommand{\U}{\operatorname{U}}

\def\sla#1{#1\kern-1.2ex/\kern.2ex}

\def\eg{e.g.~}
\def\ie{i.e.~}
\def\cf{cf.~}

\newtheorem{thm}{\bf Theorem}[section]
\newtheorem{dfn}[thm]{\bf Definition}
\newtheorem{lemma}[thm]{\bf Lemma}
\newtheorem{prop}[thm]{\bf Proposition}
\newtheorem{cor}[thm]{\bf Corollary}

\numberwithin{equation}{section}

\newcounter{mylc}
\renewcommand{\themylc}{\roman{mylc}}
\newenvironment{mylist}{\begin{list}{\themylc )}
{\usecounter{mylc}\settowidth{\labelwidth}{iiii)}}}{\end{list}}

\title{\bf Quasi-Exactly Solvable Spin \boldmath$1/2$\\
       \bf\sch\ Operators\thanks{Supported in part by DGICYT Grant
PB92--0197.}}
\author{Federico Finkel\\Artemio Gonz\'alez-L\'opez\\Miguel A.
Rodr\'\i guez\\
\\\em Departamento de F\'\i sica Te\'orica II\\\em Universidad
Complutense\\
\em 28040 Madrid, SPAIN}

\date{August 3, 1995}

\begin{document}
\maketitle
\begin{abstract}
The algebraic structures underlying quasi-exact solvability for spin
$1/2$ Hamiltonians in one dimension are studied in detail. Necessary and
sufficient conditions for a matrix second-order differential operator
preserving a space of wave functions with polynomial components to be
equivalent to a \sch\ operator are found. Systematic simplifications of
these conditions are analyzed, and are then applied to the construction
of several new examples of multi-parameter QES spin $1/2$ Hamiltonians in
one dimension.
\end{abstract}
\newpage
\section{Introduction}\label{sec.intro}

Symmetries have traditionally played an essential role in quantum
mechanics. For a few remarkable Hamiltonians, the knowledge of enough
symmetries leads to a complete characterization of the spectrum by
algebraic methods, \cite{OlPe83}. In general, however, the spectrum of an
arbitrary Hamiltonian cannot be calculated analytically. During the last
decade, a remarkable intermediate class of {\em \qes} (QES) spectral
problems was introduced, for which a finite part of the spectrum can be
computed by purely algebraic methods, \cite{Tu88}, \cite{Us89},
\cite{Sh89}. The key feature in the latter class of spectral problems is
that the Hamiltonian $H$ is expressible as a quadratic combination of the
generators of a
\fd\ Lie algebra $\g$ of first order \dfo s preserving a
\fd\ module of smooth functions $\n$. Thus, $H$ restricts to a linear
transformation in the finite-dimensional vector space $\n$, and therefore
part of its spectrum can be computed by matrix eigenvalue methods.
Appropriate boundary conditions must be imposed so that the
eigenfunctions thus obtained qualify as physical wave functions, as \eg
square integrability if they represent bound states of the system,
\cite{GKO93bis}.

These ideas, originally introduced for scalar Hamiltonians describing
spinless particles, can be generalized to include particles with spin.
The first step in this direction was taken by Shifman and Turbiner,
\cite{ShTu89}, using the fact that a Hamiltonian for a spin $1/2$
particle in $d$ spatial dimensions can be constructed from a Lie
superalgebra of first order \dfo s in
$d$ ordinary (commuting) variables and one Grassmann (anticommuting)
variable. Alternatively, \cite{BrKo93}, $2 \times 2$ matrices (or
$N\times N$ matrices for particles of arbitrary spin, \cite{BGGK94}) can
be used to represent the Grassmann variable. However, in stark contrast
with the scalar case, very few examples of matrix QES \sch\ operators
have been found thus far,
\cite{ShTu89}. There are two important conceptual reasons for this fact.
First, the algebraic structures underlying partial integrability in the
matrix case are richer and less understood than in the scalar case. For
one thing, as mentioned before, for matrix Hamiltonians Lie superalgebras
of matrix differential operators naturally come into play, whereas in the
scalar case only Lie algebras need be considered. Moreover, as we shall
explain in Section~\ref{sec.PVSP}, one even has to go beyond Lie
superalgebras of matrix differential operators in order to explain
quasi-exact solvability in the matrix case,
\cite{BrKo93}, \cite{BGGK94}. Secondly, \cite{GKO93}, \cite{GKO93bis},
every scalar second order \dfo\ in one dimension can be transformed into a
\sch\ operator of the form
$-\partial_x^2+V(x)$ by a suitable change of the independent variable $x$
and a local rescaling of the wave function. For matrix \dfo s, the
analogue of this result---$V(x)$ being now a Hermitian matrix of smooth
functions---is no longer true unless the operator satisfies quite
stringent conditions, as we shall see in detail in Section~\ref{sec.sch}.

The aim of this paper is to achieve a better theoretical understanding of
quasi-exact solvability in the matrix case, which will enable us to
construct new examples of matrix QES \sch{} operators. To this end, in
Sections
\ref{sec.scalar} and \ref{sec.PVSP} we study the algebraic properties of
certain algebras of matrix QES operators, reviewing the literature on the
subject and obtaining several new results as well. In particular, we give
a self-contained proof of the characterization of the class of QES matrix
differential operators preserving a finite-dimensional space of wave
functions with polynomial components stated by Turbiner, \cite{Tu92bis},
and Brihaye {\em et al.,} \cite{BrKo93}, \cite{BGGK94}. For the important
particular case of spin
$1/2$ particles, we derive in Section~\ref{sec.sch} necessary and
sufficient conditions for a QES operator to be equivalent to a non-trivial
\sch\ operator. These conditions turn out to be too complicated to be
solved in full generality, and so in Sections~\ref{sec.canon} and
\ref{sec.gauge} we introduce some key simplifications that will prove
very useful in the task of finding explicit examples. Finally, the
previous results are applied in Section~\ref{sec.examples} to the
construction of several new examples of multi-parameter QES spin $1/2$
Hamiltonians in one dimension.

\section{Scalar QES Operators}\label{sec.scalar}

We start with the scalar case, introducing the basic concepts and
definitions and stating two theorems for the one-dimensional case which
will play an important role in what follows. Since the results of this
section are fairly standard, we will skip many details and all the
proofs, referring the reader to the review articles \cite{Sh89} and
\cite{GKO93} for an in-depth study.

Let $M$ denote an open subset of $\Bbb R^d$, and let $\drm{M}$ be the Lie
algebra of first order \dfo s
$$
X = \sum_{i=1}^d \xi^i(z)\dd {z^i} + \eta(z),\qquad z=(z^1,\dots,z^d)\in
M,
$$
acting on $C^\infty(M)$, the Lie bracket being defined as the usual
commutator between operators:
$$
[X,Y] = X\,Y-Y\,X,\qquad X,Y\in\drm M.
$$

\begin{dfn}\label{def.qes} A \fd{} Lie subalgebra $\g$ of $\drm{M}$ is
called {\em\qes{} (QES)} if it preserves a \fd\ module $\n \subset
C^\infty(M)$. A \dfo{} $T$ is QES if it lies in the universal enveloping
algebra $\zap U(\g)$ of a QES Lie algebra $\g$.
\end{dfn}

In general, quasi-exact solvability of a given \dfo\ $T$ cannot be
ascertained {\em a priori}. Therefore, the procedure usually followed
consists in classifying QES Lie algebras modulo a suitable equivalence
relation, and then using the canonical forms in
the classification thus obtained to construct QES operators.

\begin{dfn}\label{def.equiv}
Two \dfo s $T(z)$ and $\Tb(\zb)$ are {\em equivalent} if they are
related by a change of the independent variables
\begin{equation}\label{eq.civ}
				\zb=\varphi (z)
\end{equation}
and a local scale transformation by a non-vanishing function $U(z)$, \ie
\begin{equation}\label{eq.st}
				\Tb(\bar z) = U(z)\,T(z)\,U^{-1}(z).
\end{equation}
\end{dfn}

The corresponding notion of equivalence for QES algebras follows
directly, \ie two QES Lie algebras $\g$ and $\gb$ are {\em equivalent} if
their elements can be mapped into each other by a {\em fixed}
transformation
\eqref{eq.civ}--\eqref{eq.st}. Their associated \fd\ modules $\n$ and
$\nb$ are then related by
\begin{equation}
				\nb=U \cdot \n,
\end{equation}
the functions being expressed in the appropriate coordinates. The local
classification of \fd\ QES Lie algebras under the above notion of
equivalence has already been completed for the case of one and two (real
or complex) variables. Here we shall need only the one-dimensional case,
\cite{Mi68}, \cite{Tu88}, \cite{KaOl90}, \cite{GKO91}.
\begin{thm}\label{thm.KaOl}
Every (non-singular) QES Lie algebra in one (real or complex) variable is
locally equivalent to a subalgebra of one of the Lie algebras
\begin{equation}\label{eq.gn}
		\g_n=\Span\{\Dz ,z \Dz ,z^2 \Dz - n z,1\} ,
\end{equation}
where $n\in\Bbb{N}$. The associated $\g_n$-module is $\n_n=\p{n}$,
the space of polynomials of degree at most $n$.
\end{thm}
(The two-dimensional case, which is considerably more complicated but is
not needed for the sequel, is discussed in \cite{GKO92}, \cite{GKO91}, and
\cite{GKO95}.)

According to the previous theorem, every one-dimensional (scalar) QES
differential operator $\Tb$ is locally equivalent to an operator
$T\in\zap U(\g_n)$ preserving $\p{n}$ for a suitable $n$. A partial
converse of the latter result follows from the following remarkable
theorem due to Turbiner,
\cite{Tu92}:
\begin{thm}\label{thm.Tu}
Let $\Tk$ be a $k$-th order linear \dfo{} preserving $\p{n}$. We then
have:
\begin{mylist}
\item If $n \geq k$, then $\Tk$ may be represented by a $k$-th degree
polynomial
      in the operators
\begin{equation}\label{eq.J's}
\Jp n=z^2 \Dz - n z\: ,\qquad \J0 n=z \Dz - \frac n2\: ,\qquad \Jm =\Dz\:,
\end{equation}

\item If $k>n$, then $\Tk=T\,\Dz^{n+1}+\tilde T$, where $T$ is a linear
      \dfo{} of order $k-n-1$, and $\tilde T$ is a linear differential
operator of order at most $n$ satisfying i).
\end{mylist}
\end{thm}
The operators $\{\Jp n,\J0 n,\Jm\}$ defined above span a QES Lie algebra
$\hg_n$ isomorphic to $\sl2$, and the Lie algebras $\g_n$ in
Theorem~\ref{thm.KaOl} are simply a central extension by the constant
functions of the corresponding
$\hg_n$.

\section{Algebraic Properties of PVSP Operators}\label{sec.PVSP}

In the last section we have seen that every scalar QES scalar \dfo{} in
one variable is essentially (up to equivalence) a polynomial in the
generators of a Lie algebra $\hg_n$ preserving $\p{n}$ (for suitable
$n$). When working with vector-valued wave functions, the natural
generalization of $\p{n}$ is the {\em polynomial vector space}
$\p{n_1,\dots,n_N}=\p{n_1}\oplus\dots\oplus\p{n_N}$, with elements
$\Psi(z)=(\psi_1(z),\dots,\psi_N(z))^t$ such that each component $\psi_i$
is a polynomial of degree at most $n_i$ with complex coefficients.

\begin{dfn} A $N\times N$ matrix \dfo{} $T$ is called {\em polynomial
vector space preserving (PVSP)} if it preserves $\p{n_1,\dots ,n_N}=
\p{n_1}\oplus\dots\oplus\p{n_N}$ for some non-negative integers
$n_i,\; i=1,\dots ,N$.
\end{dfn}

We will denote by $\pk{n_1,\dots ,n_N}$ the complex vector space of linear
PVSP operators of order at most $k$ preserving $\p{n_1,\dots ,n_N}$.
Following \cite{BrKo93} and \cite{BGGK94}, we will restrict ourselves in
this paper to studying matrix PVSP \dfo s. As we will be mainly concerned
with spin $1/2$ particles, the case $N=2$ deserves special attention.

\subsection{Case $N=2$}
Let $n\geq\De$ be non-negative integers, and consider the following set of
matrix \dfo s

\begin{gather}\label{eq.TJQ}
\Tp = \bpm\Jp{n-\De} & 0 \\ 0 & \Jp{n}\epm ,\quad
\T0 = \bpm\J0{n-\De} & 0 \\ 0 & \J0{n}\epm ,\quad
\Tm = \bpm\Jm & 0 \\ 0 & \Jm\epm ,\notag \\[.5cm]
J=\frac{1}{2} \bpm n+\De & 0 \\ 0 & n\epm ,\\[.5cm]
\qquad\Qa = z^\al\,\sigma^- ,\qquad \Qab = \qab{n}{\De}\,\sigma^+ \,
,\quad
\al=0,\dots ,\De,\notag
\end{gather}
with
\begin{equation}
\qab{n}{\De}=\prod_{k=1}^{\De-\al} (z\Dz -n+\De-k) \,\Dz^\al\,,
\end{equation}
where we have adopted the convention that a product with its lower limit
greater than the upper one is automatically $1$, and
$\sigma^+=(\sigma^-)^t=\bpm 0 & 1\\0 & 0\epm$.\\

It can be easily checked that the $6+2\De$ operators in~\eqref{eq.TJQ}
(and also any polynomial thereof) preserve $\p{n-\De,n}$. We now
introduce a $\Bbb{Z}_2$-grading in the set of
$2\times 2$ matrix \dfo s $\zap{D}_{2\times 2}$ as follows: an operator
$$T=\bpm a & b\\c & d\epm ,\qquad a,\: b,\: c,\: d \in\zap{D},$$ is said
to be {\em even} if $b=c=0$, and {\em odd} if $a=d=0$. Therefore, the
$T$'s and $J$ are even and the $Q$'s and $\Qb$'s odd. This grading,
combined with the usual product (composition) of operators, endows
$\zap{D}_{2\times 2}$ with an associative superalgebra structure. We can
also construct a Lie superalgebra in
$\zap{D}_{2\times 2}$ by defining a generalized Lie product by
\begin{equation}\label{eq.prod}
[A,B]_s=AB-(-1)^{\deg A \deg B}BA.
\end{equation}
However, this product does not close within the vector space spanned by
our operators
\eqref{eq.TJQ}, except for $\De=0,1$. The explicit commutation relations
are as follows, \cite{BrKo93}, \cite{BGGK94}:

\begin{align}\label{eq.comm}
& [\Tp,\Tm] = -2\T0 ,\hspace{1cm} [T^\pm,\T0] = \mp T^\pm ,\notag\\[.3cm]
& [J,\Te] = 0 ,\hspace{1cm}
          [J,\Qa] = -\frac{\De}{2}\Qa ,\hspace{1cm}
                   [J,\Qab] = \frac{\De}{2}\Qab ,\notag\\[.3cm]
& [\Qa,\Te] = \big(-\al+\frac{\De}{2}(1+\ep)\big) Q_{\al+\ep}
,\hspace{1cm}
       [\Qab,\Te]= \big(\al-\frac{\De}{2}(1-\ep)\big)
\Qb_{\al-\ep},\notag\\[.3cm] & \{\Qab,Q_\be\} = \begin{cases}
                     \M\, (\Tm)^{\al-\be}\,, & \quad\al\geq\be\\
                     (\Tp)^{\be-\al}\, M_{\be\al}\,, & \quad\be\geq\al,
                   \end{cases}\notag\\[.3cm]
& \{\Qa,Q_\be\}=\{\Qab,\Qb_\be\}=0,
\end{align}
where $\ep = +,0,-$, and $\M$ is given, for $\al\geq\be$, by
$$
\M=\prod_{j=0}^{\De-\al-1}(\T0+J_c-j-\be P_2)
   \prod_{k=0}^{\be-1}(\T0+J-k-(\De-\al) P_1)
$$
with $J_c=\De-1-J$, and $P_1=1-P_2=\bpm 1 & 0\\0 & 0\epm$.\\

As shown in \cite{BrKo93}, $\M$ can be expressed in terms of $\T0$, $J$,
the identity, and the Casimir (for the even subalgebra)
\begin{equation}\label{eq.Cas}
C=-\frac{1}{2}(\Tp\Tm+\Tm\Tp)+\T0\T0
   =\frac{1}{4}\bpm m(m+2) & 0\\0 & n(n+2)\epm,
\end{equation}
where $m=n-\De$, independently from $n$ and the projectors $P_1$ and
$P_2$. It can be readily verified that $\{\Qab,Q_\be\}$ gives a $\De$-th
order even \dfo, so the vector space spanned by the operators in
\eqref{eq.TJQ} is not closed under the Lie product \eqref{eq.prod}
whenever $\De\geq2$. Moreover, it is not difficult to show that the Lie
superalgebra $\s_\De$ generated by the operators \eqref{eq.TJQ} is in
this case infinite dimensional. Indeed, if we commute
$\{\Qb_{\De},Q_{0}\}=(\Tm)^\De$ with $\{\Qb_{0},Q_{\De}\}=(\Tp)^\De$
iteratively we obtain monomials in $\Tp,\:\T0,\:\Tm$ of increasingly
higher order. For $\De=1$ the underlying algebraic structure is the
classical simple Lie superalgebra $\frak{osp}(2,2)$, \cite{Sh89},
\cite{Tu92bis}, whereas for $\De=0$ it is $\frak{h}_1\oplus\sl2$, where
$\frak{h}_1$ is the \mbox{3-di}mensional Heisenberg superalgebra. As
remarked in \cite{BrKo93}, in this latter case we can leave the grading
aside and replace $J=1$ by $\tilde J=\sigma_3$, ending up with the Lie
algebra $\sl2\oplus\sl2$.

Our next objective is to prove the analogous of Theorem~\ref{thm.Tu} for
PVSP operators preserving $\p{n-\De,n}$ (first mentioned without proof by
Turbiner for $\De=1$ in \cite{Tu92bis}, and subsequently by Brihaye and
Kosinski for arbitrary $\De$, \cite{BrKo93}). It turns out that the
operators \eqref{eq.TJQ} play the same role for the matrix case as the
$J$'s in \eqref{eq.J's} do for the scalar one. The proof relies on the
next two Lemmas.

\begin{lemma}\label{lem.down}
Let $\Tk:\p{n}\rightarrow\p{n-\De}$ be a $k$-th order linear \dfo{}, with
$n\geq\De$. We then have:
\begin{mylist}
\item If $n\geq k\geq\De$, then $\Tk$ may be represented as
\begin{equation}\label{eq.down}
\Tk=\sum_{\al=0}^\De \qab{n}{\De}\, p_{k-\De}^\al(\Je{n}),
\end{equation}
where $p_{k-\De}^\al(\Je{n})$ are polynomials of degree not higher than
$k-\De$ in the operators $\Jp{n},\J0{n}$ and $\Jm$.
\item If $\De>k$, then $\Tk=0$.
\item If $k>n$, then $\Tk=T\,\Dz^{n+1}+\tilde T$, where $T$ is a linear
      \dfo{} of order $k-n-1$, and $\tilde T$ is a linear differential
      operator of order at most $n$ satisfying i) or ii).
\end{mylist}
\end{lemma}

{\noindent\em Proof}. {\em i).} The essential point of the argument is
that, as $\p{n-\De}\subset\p{n}$, $\Tk$ preserves also $\p{n}$, and so it
must be a $k$-th degree polynomial in $\Je{n}$ with suitable additional
restrictions on its coefficients, according to Theorem~\ref{thm.Tu}. We
will prove that
\begin{align}\label{eq.down2}
\Tk=\, & \qb{0}{n}{\De}\,p_{k-\De}^0(\Jp{n},\J0{n})\,+\,
\sum_{\al=1}^{\De}\qab{n}{\De}\,p_{k-\De}^\al(\J0{n})\notag\\
& +\qb{\De}{n}{\De}\,\Jm\,p_{k-\De-1}(\J0{n},\Jm),
\end{align}
which clearly implies \eqref{eq.down}.

We will proceed by induction on $\De$. First assume $\De=1$. Without any
loss of generality (because of the scalar Casimir analogous to
\eqref{eq.Cas}) we may write
$$
\Tk_1=\sum_{r+s\leq k}c_{rs}(\J0{n})^r(\Jp{n})^s\,+\,
\Jm\,p_{k-1}^1(\J0{n})\,+\,
(\Jm)^2\,p_{k-2}(\J0{n},\Jm),
$$
for some constants $c_{rs}$. As $\Jp{n}$ annihilates $z^n$, acting with
$\Tk_1$ on $z^{n-s}$, $s=0,\dots,k-1$, we find
$$
\Tk_1\,z^{n-s}=
(-1)^s s!\,\sum_{r=0}^{k-s}\left(\frac{n}{2}\right)^r c_{rs}\,z^n\,+\,
\text{lower order terms,}
$$
leading to $\sum_{r=0}^{k-s}(\frac{n}{2})^r c_{rs}=0$ for each $s$.
It can be easily shown that the most general element in
$\Span\{(\J0{n})^r(\Jp{n})^s\}_{r=0}^{k-s}\,$ satisfying this condition is
$$
(\J0{n}-\frac{n}{2})
\sum_{r=0}^{k-s-1}\tilde{c}_{rs}\,(\J0{n})^r(\Jp{n})^s.
$$
Finally, no monomial $(\Jp{n})^k$ may be present in $\Tk_1$. Thus,
$$
\Tk_1=\qb{0}{n}{1}\,p_{k-1}^0(\Jp{n},\J0{n})\,+\,
\qb{1}{n}{1}\,\big( p_{k-1}^1(\J0{n})\,+\,
\Jm\,p_{k-2}(\J0{n},\Jm)\big),
$$
completing the proof of \eqref{eq.down2} for $\De=1$. Now assume that
\eqref{eq.down2} holds for every $\Tk_{\De}:\p{n}\rightarrow\p{n-\De}$.
In particular, it is also true for an operator $\Tk_{\De+1}$ mapping
$\p{n}$ into
$\p{n-\De-1}$. As $\qab{n}{\De}$ ($\al=0,\dots,\De-1$) annihilates
$z^{n-s}$ if $0\leq s<\De-\al$, acting with $\Tk_{\De+1}$ on $z^{n-s}$
($s=0,\dots,k$) and reasoning as in the case $\De=1$, we deduce that a
factor $(\J0{n}-\frac{n-\De+\al}{2})$ must be present in front of each
$p_{k-\De}^\al$ for $\al=0,\dots,\De$. If we move these factors before
their respective $\qab{n}{\De}$ and group together the monomials in
$p_{k-\De-1}(\J0{n},\Jm)$ with no $\Jm$, we find that $\Tk_{\De+1}$ may be
written in the form \eqref{eq.down2} with $\De$ replaced by $\De+1$.\\

{\em ii).} If we apply {\em i)} for $\De=k$, we find
$\Tk=\sum_{\al=0}^{k}c_\al\,\qab{n}{k}$, and acting on $z^{n-s}$,
$s=0,\dots,k$,  we conclude that $c_\al=0$ for all $\al$.\\

{\em iii).} This is obvious, since derivatives of order higher than $n$
annihilate
$\p{n}$.\QED\\

The corresponding case $\Tk:\p{n}\rightarrow\p{n+\De}$ is treated next:

\begin{lemma}\label{lem.up}
Let $\Tk:\p{n}\rightarrow\p{n+\De}$ be a $k$-th order linear \dfo{}.
We then have:
\begin{mylist}
\item If $n\geq k$, then $\Tk$ may be represented as
$$
\Tk=\sum_{\al=0}^\De z^\al\, p_k^\al(\Je{n}),
$$
where each $p_k^\al$ is a polynomial of degree not higher than $k$ in
the operators $\Jp{n},\J0{n}$ and $\Jm$.

\item If $k>n$, then $\Tk=T\,\Dz^{n+1}+\tilde T$, where $T$ is a linear
      \dfo{} of order $k-n-1$, and $\tilde T$ is a linear differential
      operator of order at most $n$ satisfying i).
\end{mylist}
\end{lemma}

The proof of this Lemma is similar to that of Theorem~\ref{thm.Tu}, and
shall not be presented here. The Theorem analogous to \ref{thm.Tu} for a
PVSP operator in $\pk{m,n}$ now follows applying Theorem~\ref{thm.Tu} and
the Lemmas above to each entry of the operator:

\begin{thm}\label{thm.TBK}
Let $n\geq m$, and $\De=n-m$. Let $\Tk$ be $k$-th order \dfo{} in
$\pk{m,n}$. We then have:
\begin{mylist}
\item If $m\geq k$, then $\Tk$ is a polynomial in the
operators~\eqref{eq.TJQ}. More explicitly, if $k\geq\De\geq 1$,
\begin{multline}\label{eq.TBK}
\Tk =p_k(\Te)+J\,\tilde p_k(\Te)+\sum_{\al=0}^\De
\Qab\,\pb_{k-\De}^\al(\Te)+
     \sum_{\al=0}^\De \Qa\, p_k^\al(\Te)\,,
\end{multline}
while if $\De=0$ the same formula is valid with $J$ replaced by $\tilde
J$. If $k<\De$, every $\pb_{k-\De}^\al$ must be identically zero.

\item If $n\geq k>m$, then $\Tk=T\,\Dz^{m+1}+\tilde T$, where
$T$ and $\tilde T$ are matrix linear \dfo s of the form
$$
T=\bpm a^{(k-m-1)} & 0\\c^{(k-m-1)} & 0\epm,\qquad
\tilde T=\bpm \tilde a^{(m)} & \tilde b^{(k)}\\
              \tilde c^{(m)} & \tilde d^{(k)}\epm,
$$
where the superscripts indicate the highest possible derivative in
each entry, and $\tilde T$ satisfies i).

\item If $k>n$, then $\Tk=T\,\Dz^{n+1}+\tilde T$, where $T$ is a $2\times
2$ matrix linear \dfo{} of order $k-n-1$, and $\tilde
T:\p{m,n}\rightarrow\p{m,n}$ is a linear PVSP operator of order at most
$n$ verifying i) or ii).
\end{mylist}
\end{thm}

The next issue to be addressed is to find out the number of parameters
determining a generic $k$-th order linear \dfo{} preserving $\p{m,n}$,
that is, the dimension of $\pk{m,n}$. In the scalar case, any $k$-th
degree polynomial in
$\Jp{n},\:\J0{n},\:\Jm$ may be constructed from the monomials
$\{(J^\pm_n)^r(\J0{n})^{s-r}\}_{r=0}^s,\; s=0,\dots,k$, and so
$\dim \pk{n}=(k+1)^2$ if $n\geq k$, \cite{Tu92}.
Remarkably, in the matrix case we have $\dim \pk{m,n}=4(k+1)^2$
independently of $m$ and $n$, provided $m\geq k\geq n-m$, \cite{BrKo93},
as a consequence of the following Lemma:

\begin{lemma}\label{lem.basis}
The following set of monomials form a basis of the vector space of
polynomials in the operators~\eqref{eq.TJQ} of differential order at most
$k$:
$$
\{X(T^\pm)^r(\T0)^{s-r}\}_{r=0}^s \,,\; \{\Qa(\Tp)^s\}_{\al=1}^\De, \quad
s=0,\dots,k,
$$
where $X=1$, $J$ (or $\tilde J$, if $\De=0$), $Q_0$, along with, if
$k\geq\De$:
$$
\{\Qb_0(T^\pm)^r(\T0)^{s-r}\}_{r=0}^s \,,\; \{\Qab(\Tm)^s\}_{\al=1}^\De,
\quad s=0,\dots,k-\De.
$$
\end{lemma}

{\noindent\em Proof}. Linear independence of the monomials is
straightforward from the definition of the operators. Completeness  is a
consequence of the following facts. In the first place, every $J^s$ is a
linear combination of $\{1,J\}$ (and analogously for $\tilde J$).
Secondly,
$J\Qa$ is proportional to $\Qa$, and $\Qa Q_\be=0$ (and the same for the
$\Qb$'s). Third, any product $\Qa\Qb_\be$ is a diagonal PVSP operator,
and thus expressible through the $T$'s and $J$ (or $\tilde J$). Finally,
the formulas ($\al\geq 1$)
\begin{xalignat*}{2}
\Qa \T0 & = Q_{\al-1}\Tp+\frac{n-\De}{2}\,\Qa\,, & \qquad
\Qa \Tm & = Q_{\al-1}\T0+\frac{n-\De}{2}\,Q_{\al-1}\,,\\
\Qab \T0 & = \Qb_{\al-1}\Tm+\big(\frac{n}{2}+1\big)\,\Qab\,, & \qquad
\Qab \Tp & = \Qb_{\al-1}\T0+\big(\frac{n}{2}+1\big)\,\Qb_{\al-1}\,,
\end{xalignat*}
allow us to remove every $\T0$ and $\Tm$ (respectively $\Tp$) from the
monomials with $\Qa$ (respectively $\Qab$), $\al=1,\dots,\De$. \QED\\

\begin{cor}\label{cor.par}
Let $n\geq m\geq k$, and $\De=n-m$. We then have:
$$
\dim\pk{m,n}= \begin{cases}
                 4(k+1)^2\,, & \quad k\geq\De \\
                 (k+1)(3k+\De+3)\,,& \quad \De>k.
              \end{cases}
$$
\end{cor}
If $m<k$, $\dim \pk{m,n}$ is no longer finite, as arbitrary \dfo s are
involved in this case.

\subsection{Case $N>2$}
We now examine briefly some aspects of the case $N>2$.  Let
$n_1\leq\dots\leq n_N$ be non-negative integers, and let
$\De_{ij}=n_j-n_i$, where $j>i$. Consider the following set of $N\times
N$ matrix differential operators,
\cite{BGGK94}:
\begin{align}\label{eq.TPQ}
& \Te=\diag(\Je{n_1},\dots,\Je{n_N})\,,\qquad\ep = +,0,-,\notag\\[.3cm]
& P_{i}=\diag(0,\dots,0,1,0,\dots,0)\,,\notag\\[.3cm]
& \Qa(i,j)=z^\al\,\lambda_{ij}\,,\qquad i>j\,,\qquad\al=0,\dots,\De_{ji},
\notag\\[.3cm]
& \Qab(i,j)=\qab{n_j}{\De_{ij}}\,\lambda_{ij}\,,\qquad j>i\,,\qquad
\al=0,\dots,\De_{ij},
\end{align}
where $(\lambda_{ij})_{pq}=\delta_{ip}\delta_{jq}$. It can be readily
verified that the operators in~\eqref{eq.TPQ} preserve
$\p{n_1,\dots,n_N}$. A complication arising when $N>2$ is to define a
suitable composition law between the latter operators. In the approach of
Brihaye {\em et al.}, \cite{BGGK94}, this composition law is defined to
be an anticommutator if both operators are off-diagonal and a commutator
otherwise, but the algebra thus obtained is no longer a Lie superalgebra,
since the anticommutator of two off-diagonal operators is not always a
diagonal one. This reflects the fact that the $\Bbb{Z}_2$-grading we
introduced for $N=2$ (\ie classifying the operators in diagonal and
off-diagonal) does not define an associative superalgebra in
$\zap{D}_{N\times N}$ when $N>2$, for the usual product (composition) of
two off-diagonal matrix \dfo s is not necessarily diagonal. A possible
generalization of this $\Bbb{Z}_2$-grading, endowing $\zap{D}_{N\times
N}$ with an associative superalgebra structure can be defined as follows.
An operator $T=a\lambda_{ij}$, where $a\in\zap{D}$, is said to be even
(respectively odd) if $i+j$ is even (respectively odd). Hence, any
diagonal operator is even. We can likewise use this grading and the
generalized Lie product~\eqref{eq.prod} to construct a Lie superalgebra
structure in $\zap{D}_{N\times N}$. It is not clear, however, whether
this construction is really useful, and so it will not be further
discussed.

As remarked by Brihaye {\em et al.}, \cite{BGGK94}, Theorem~\ref{thm.TBK}
can be easily generalized to arbitrary $N$, the operators \eqref{eq.TPQ}
playing the same role as those in~\eqref{eq.TJQ} for $N=2$. Moreover, it
is not difficult to show that
$\dim \pk{n_1,\dots,n_N}$ is still independent of the $n_i$'s if they are
large enough and their differences are small enough. More precisely, if
$n_1\geq k$, we have:
$$
\dim \pk{n_1,\dots,n_N}=
\begin{cases}
   N^2(k+1)^2\,, & \quad k>\De_{1N}\\[.3cm]
   {\displaystyle\frac{N(N+1)}{2}
      (k+1)^2\,+\,(k+1)\,\sum_{i<j}\theta_{ij}}\,,& \quad \De_{1N}>k\,,
\end{cases}\vspace{.2cm}
$$
where $\theta_{ij}=\De_{ij}$ if $\De_{ij}>k$, and $\theta_{ij}=k+1$
if $\De_{ij}\leq k$. If $k>n_1$, arbitrary \dfo s are involved and
thus $\pk{n_1,\dots,n_N}$ is infinite-dimensional.

Although no attempt will be made here to give a formal definition
of a QES algebra of matrix \dfo s, it is clear that
Definition~\ref{def.qes} of a QES \dfo{} is too restrictive in the matrix case.
Indeed, the results of this section suggest that in the matrix case
one should include at least Lie superalgebras of \dfo s---not necessarily
\fd{} nor spanned by first order operators---preserving a \fd{} module of
functions among the class of matrix QES algebras. In any case, it
is intuitively clear that PVSP operators are just a particular class of
QES operators.

\section{Spin \boldmath$1/2$ \sch\ Operators}\label{sec.sch}

{}From now on we will deal only with $2\times 2$ matrix second order \dfo s
($N=k=2$ in the notation of the previous sections). We start by formally
defining the class of matrix \sch{} operators:

\begin{dfn}
A {\em \sch-like operator} is a second order \dfo{} of the form
$H=-\Dx^2+V(x)$, where $V$ is an arbitrary $2\times 2$ (complex) matrix.
A {\em\sch{} operator} (or {\em Hamiltonian}) is a Hermitian \sch-like
operator, \ie the matrix $V$ is of the form
\begin{equation}\label{eq.V}
V=\bpm v_1(x) & v^\ast(x) \\ v(x) & v_2(x)\epm\,,
\end{equation}
where $v_1$ and $v_2$ are real-valued functions and $v$ is an arbitrary
complex-valued function.
\end{dfn}

The notion of equivalence we shall use for matrix differential operators
is the same as in the scalar case (see Definition~\ref{def.equiv}), where
now the gauge factor $U(z)$ is an invertible complex $2\times 2$ matrix.
We will be interested in constructing one-dimensional \sch{} operators
$H$ equivalent to a second order \dfo{} $T$ in $\pp{m,n}$, with
$\De=n-m\geq 0$. This equivalence can then be used to construct
$m+n+2$ eigenfunctions of $H$ from the corresponding ones of $T$ obtained
by diagonalization of the $(m+n+2)\times(m+n+2)$ Hermitian matrix
representing $T$ in $\p{m,n}$. We will assume that $m\geq 2$, and thus
$T$ is a polynomial in the operators \eqref{eq.TJQ}, according to
Theorem~\ref{thm.TBK}.

Let $T:\p{m,n}\rightarrow\p{m,n}$ ($n\geq m\geq 2$) be a second-order PVSP
operator. From Theorem~\ref{thm.TBK} we have
\begin{equation}
\label{eq.T}
-T=A_2(z)\,\Dz^2+A_1(z)\,\Dz+A_0(z)\,,
\end{equation}
where the $A_i$'s are $2\times2$ matrices with polynomial entries (in
this section, capital letters will be reserved for matrices). Assume that
$T(z)$ is equivalent to a \sch{} operator $H(x)$ under a gauge
transformation
$U(z)$ and a local change of variable by a real-valued function
$x=\varphi (z)$, \ie
\begin{align}\label{eq.equiv1}
-T(z) & =-U^{-1}(z)H(x)U(z)\notag\\
      & =\Dx^2+2A\,\Dx-B,
\end{align}
with
$$
A(x)=\Ut^{-1}\Ut_x\,,\qquad B(x)=\Ut^{-1}V\Ut-A^2-A_x\,,\qquad
\Ut(x)=U(\varphi^{-1}(x))\,.
$$
Here and in what follows, a subscripted $x$ denotes derivation with
respect to
$x$, while derivatives with respect to $z$ will be denoted with a prime
$'$. Expressing $T(z)$ in the variable $x$, we obtain the operator
$\tilde T(x)$ given by
\begin{equation}\label{eq.equiv2}
-\tilde T(x)= \bigl[A_2
{\varphi'}^2\,\Dx^2+(A_1\varphi'+A_2\,\varphi'')\,\Dx+A_0
              \bigr]_{\varphi^{-1}(x)}\,,
\end{equation}
and comparing with \eqref{eq.equiv1} we conclude that $A_2$ must be a
multiple of the identity. It then follows that only the first term in
equation~\eqref{eq.TBK} of Theorem~\ref{thm.TBK} contributes to $A_2$, and
taking into account the explicit form of the $\Te$'s (see
\eqref{eq.TJQ}), we conclude that $A_2$ is a $4$-th degree polynomial
$p_4$ times the identity matrix.  (Unless otherwise stated, we will
denote by $p_n(z)$ an arbitrary polynomial in
$z$ of degree at most $n$ with complex coefficients.) We also deduce that
$\varphi(z)$ satisfies the equation $p_4 {\varphi'}^2=1$, or
\begin{equation}\label{eq.xofz} x=\int^z\frac{1}{\sqrt{p_4(s)}}\,ds\,,
\end{equation} and thus the coefficients of $p_4$ must be real.
Identifying the corresponding remaining terms in \eqref{eq.equiv1} and
\eqref{eq.equiv2}, we then get
\begin{equation}\label{eq.AB}
A(x)=\frac{1}{2\sqrt{p_4}}\big(A_1-\frac{1}{2}p_4'\big)\big|_{\varphi^{-1}(x)}\,,
\qquad B(x)=-A_0\big|_{\varphi^{-1}(x)}\,.
\end{equation}
Thus, we have shown:

\begin{thm}
Let $T$ be a PVSP operator in $\pp{m,n}$, with $n\geq m\geq 2$.
Then $T$ is equivalent to a \sch-like operator if and only if it is
of the form
\begin{equation}\label{eq.T2}
-T=p_4\,\Dz^2+A_1\,\Dz+A_0\,.
\end{equation}
The operator $T$ is equivalent to a \sch{} operator $-\Dx^2+V(x)$
if and only if~\eqref{eq.T2} holds, and in addition there is an invertible
matrix
$\Ut$ satisfying the differential equation
\begin{equation}\label{eq.Ux}
\Ut_x=\Ut A
\end{equation}
and such that
\begin{equation}\label{eq.hermitx}
V=\Ut\Wt\Ut^{-1}\,,\qquad\text{where}\qquad\Wt=B+A^2+A_x\,,
\end{equation}
is Hermitian (with $x$, $A$, and $B$ given by \eqref{eq.xofz} and
\eqref{eq.AB}).
\end{thm}

The eigenfunctions of the Hamiltonian $H$ are of the form
$\PSI(x)=\Ut \tilde\Psi$ with $\tilde\Psi(x)=\Psi(\varphi^{-1}(x))$, where
$\Psi(z)$ is an eigenfunction of $T$ and $\PSI$ must satisfy suitable
boundary conditions to qualify as a physical wave function.

We note that once an invertible solution $\Ut$ of equation~\eqref{eq.Ux}
has been found, any other invertible solution is of the form $U_0\Ut$ for
some $U_0$ in $\GL(2,\Bbb C)$. In fact, multiplying $\Ut$ by such $U_0$
is equivalent to performing a further constant scale transformation by
$U_0$. In the scalar case this additional freedom is absent, for \dfo s
are unaffected by scale transformations by constant functions. Note also
that a scale transformation by an arbitrary constant matrix
$U_0$ will not map every Hamiltonian into another Hamiltonian, unless
$U_0$ belongs to
$\Bbb R^+\times\U(2)$.

A matrix \dfo{} will be called {\em uncoupled} if it is either upper or
lower triangular. Since $A_2$ in \eqref{eq.T} must be a multiple of the
identity matrix, it follows that a PVSP operator of the form
\eqref{eq.T2} will be automatically uncoupled whenever
$\De>1$, as no
$\Qb$ can then be present in~\eqref{eq.TBK}. Moreover, the following
result shows that any Hamiltonian we may obtain from $T$ when $\De>1$
will be essentially diagonal:

\begin{prop}
Every Hamiltonian $H$ obtained from an uncoupled PVSP operator $T$ of the
form~\eqref{eq.T2} is diagonal, up to equivalence.
\end{prop}

{\noindent\em Proof}. If $T$ is uncoupled, the integration of equation
\eqref{eq.Ux} is straightforward. Multiplying $\Ut$ from the left by an
appropriate $U_0$ in $\Bbb R^+\times\U(2)$, we construct a new gauge
factor uncoupled in the same way as $T$. Using this new gauge factor, we
obtain  a Hamiltonian $\hat H$ given by
$$\hat H=U_0HU_0^{-1}=U_0\Ut\tilde T(U_0\Ut)^{-1}\,,$$ which is both
diagonal and equivalent to the initial one.\QED\\

Consequently, we shall limit ourselves in what follows to the cases
$\De=0,1$.

There are two main difficulties associated with the method just outlined
for constructing QES spin $1/2$ Hamiltonians. In the first place, one
needs to invert the elliptic integral~\eqref{eq.xofz} in order to compute
$z$ as a function of
$x$, which is no easy task. Secondly, the differential equation
\eqref{eq.Ux} cannot in general be solved in closed form, thus preventing
us from verifying the Hermiticity of $V$. The former complication can be
overcome, as we shall see in the next section. The latter is more
difficult to handle, although imposing further constraints on the initial
PVSP operator will contribute to simplify the problem, as shown in
Section~\ref{sec.gauge}.

We shall finish this section with a few remarks on the physical
significance of matrix \sch{} operators. First of all, one-dimensional
$2\times2$ matrix
\sch{} operators can be obtained by separation of variables from the
three-dimensional Pauli Hamiltonian describing a spin
$1/2$ charged particle in non-relativistic quantum mechanics. Consider,
indeed, the Pauli Hamiltonian
$$
H_{\mathrm{Pauli}} = (i\nabla+e\,\mb A)^2+e\,\phi-e\,\bs\sigma\cdot\mb
B\,,
$$
where $\phi$ and $\mb A=(A^1,A^2,A^3)$ are respectively the scalar and
vector potential of the external electromagnetic field, $\mb
B=\nabla\times
\mb A$ is the magnetic field, the Pauli matrices
$\bs\sigma=(\sigma^1,\sigma^2,\sigma^3)$ are given by
$$
\sigma^1=\begin{pmatrix} 0 & 1\\ 1 & 0
\end{pmatrix},\qquad
\sigma^2=\begin{pmatrix} 0 & -i\\ i & 0
\end{pmatrix},\qquad
\sigma^3=\begin{pmatrix} 1 & 0\\ 0 & -1
\end{pmatrix},
$$
$e$ is the electric charge, and physical units have been chosen so that
$\hbar=c=2m=1$. If, for example, the vector and scalar potentials depend
only on the $x$ coordinate (and we take, without loss of generality,
$A^1=0$) then
$H_{\mathrm{Pauli}}$ obviously commutes with the $y$ and $z$ components
of the linear momentum. The eigenfunctions of $H_{\mathrm{Pauli}}$ can
then be sought in the form
$$
e^{i(p_y\,y+p_z\,z)}\PSI(x)\,,
$$
where $p_y,p_z\in\Bbb R$ are the values of the $y$ and $z$ components of
the linear momentum, and the two-component spinor
$\PSI(x)$ is an eigenfunction of the one-dimensional matrix \sch{}
operator with potential \eqref{eq.V} given by
\begin{align*} v_j(x) &=  e\,\phi+(eA^2-p_y)^2+(eA^3-p_z)^2+(-1)^j
e\,\frac{dA^2}{dx}\,,\qquad j=1,2\,,\\ v(x) &= i\,e\,\frac{dA^3}{dx}\,.
\end{align*}

More surprisingly, one-dimensional $2\times2$ matrix operators are also
directly related to Dirac's relativistic equation for a spin $1/2$
charged particle in an external electromagnetic field. To see this, let us
write the latter equation as
\begin{equation}\label{eq.dirac}
(i\sla\d-e\sla A-m)\Psi(x)=0\,,
\end{equation}
where $\sla a=\gamma^\mu a_\mu$, the $\gamma^\mu$'s are
$4\times4$ matrices satisfying
$$
\left\{\gamma^\mu,\gamma^\nu\right\}=2g^{\mu\nu}\,,
$$
the metric tensor
$(g_{\mu\nu})=\diag(1,-1,-1,-1)$ is used to raise and lower indices,
$\d_\mu=\d/\d x^\mu$,
$x^0=t$,
$A^0=\phi$ and $m$ is the particle's mass. Multiplying Dirac's equation
by the operator
$i\sla\d-e\sla A+m$ we easily arrive at the second-order equation
\begin{equation}\label{eq.d2t}
\left[(i\d-eA)^2-m^2-\frac e2 F_{\mu\nu}\sigma^{\mu\nu}\right]\Psi=0\,,
\end{equation} where
$
\sigma^{\mu\nu}=\frac i2\left[\gamma^\mu,\gamma^\nu\right]
$
and $F_{\mu\nu}=\d_\mu A_\nu-\d_\nu A_\mu$ is the electromagnetic field
strength tensor. Conversely, if $\Phi$ is a solution of \eqref{eq.d2t}
then either $(i\sla\d-e\sla A+m)\Phi$ or
$\gamma_5\Phi\equiv i\gamma^0\gamma^1\gamma^2\gamma^3\Phi$ (but not both
simultaneously!) is a non-trivial solution of Dirac's equation
\eqref{eq.dirac}. In the chiral representation of the gamma matrices,
\cite{ItZu80}, we have
$$
\sigma^{0j}=i\bpm \sigma^j & 0\\ 0 & -\sigma^j\epm,\qquad
\sigma^{jk}=\epsilon_{jkl}\bpm \sigma^l & 0\\ 0 & \sigma^l\epm
$$
($j,k,l=1,2,3$, and summation over $l$ is understood), and therefore
\eqref{eq.d2t} decouples into two independent equations for the upper and
lower components
$\Psi_\pm(x)$ of $\Psi(x)$, namely (cf. \cite{FeGM58})
$$
\left[(i\d-eA)^2-m^2+e\,\bs\sigma\cdot(\mb B \mp i\,\mb
E)\right]\Psi_\pm=0\,.
$$
If the electromagnetic four-potential $A^\mu$ is time-independent, and we
look for solutions of Dirac's equation with well-defined energy $E$, \ie
we set $\Psi_\pm = e^{-iEt}\PSI_\pm(x,y,z)$, we obtain the following
equation for $\PSI_\pm$:
\begin{equation}\label{eq.d2}
\null\kern-36pt
\left[(i\nabla+e\mb A)^2-(E-eA^0)^2+m^2-e\,\bs\sigma\cdot(\mb B \mp i\,\mb
E)\right]\PSI_\pm=0\,,
\end{equation}
which has the same structure as Pauli's non-relativistic
equation. Just as was the case with Pauli's equation, separation of
variables in \eqref{eq.d2} often leads to the eigenvalue problem for a
one-dimensional matrix \sch{} operator. For instance, suppose that
$A^0=0$ and that $\mb A$ has cylindrical symmetry, that is
$$
\mb A = A(\rho)\,\mb e_z
$$
in cylindrical coordinates $(\rho,\ph,z)$. The left-hand side of
\eqref{eq.d2} then commutes with the $z$ components of the linear momentum
($-i\d_z$) and the total angular momentum ($-i\d_\ph+\frac12\sigma^3$),
which allows us to look for solutions of \eqref{eq.d2} of the form
\begin{equation}\label{eq.r12}
\PSI(\rho,\ph,z)=e^{ip_z z}\bpm R_1(\rho)e^{i(j_z-1/2)\ph}\\
R_2(\rho)e^{i(j_z+1/2)\ph}\epm.
\end{equation}
Here we have dropped the subscript $\pm$, since $\PSI_+$ and $\PSI_-$
satisfy the same equation, $p_z\in\Bbb R$ is the value of the
$z$ component of the linear momentum, and
$j_z\in\Bbb N+\frac12$ is the value of the
$z$ component of the total angular momentum. Substituting \eqref{eq.r12}
into
\eqref{eq.d2} we arrive at the following equation for $(R_1\ R_2)^t$:
\begin{multline*}
\biggl[-\d_\rho^2-\frac1\rho\d_\rho+(p_z-eA)^2+m^2-E^2\\
+ \frac1{\rho^2}\bpm (j_z-1/2)^2&0\\0&(j_z+1/2)^2\epm
+ e\frac{dA}{d\rho}\,\sigma^2\biggr]
\bpm R_1\\R_2\epm = 0\,.
\end{multline*}
Defining
$$
f_i(x) = x^{1/2} R_i(x)\,,\qquad i=1,2 \,,
$$
we finally obtain that the two-component spinor $\bigl(f_1(x)\
f_2(x)\bigr)^t$ is an eigenfunction of the one-dimensional $2\times2$
matrix
\sch{} operator with potential
$$
V(x) =
\bigl(p_z-eA(x)\bigr)^2+e\frac{dA}{dx}(x)\,\sigma^2+
\frac{j_z}{x^2}(j_z-\sigma^3)\,,
\qquad x\in (0,\infty)\,,
$$
with eigenvalue $E^2-m^2$ and boundary condition $f_1(0)=f_2(0)=0$.

\section{\boldmath$\GL(2)$ Action and Canonical Forms}\label{sec.canon}

In this section we will study how the $\GL(2)$ action on the projective
line
$\RP$ induces an automorphism in the superalgebras $\s_\De$ generated by
the operators~\eqref{eq.TJQ}. This will allow us to reduce the polynomial
$p_4(z)$ to some simple canonical forms, facilitating the evaluation of
the integral~\eqref{eq.xofz}. These ideas were first applied in the
context of QES systems to analyze the normalizability of the wave
functions of scalar QES Hamiltonians, \cite{GKO93bis}. We introduce some
definitions and results in the scalar case, and then show how to extend
these concepts to the matrix superalgebras $\s_\De$.

The action of $\GL(2)=\GL(2,\Bbb R)$ on $\RP$ via linear fractional or
M\"obius transformations,
\begin{equation}\label{eq.mobius}
z\mapsto w=\frac{\al z+\be}{\ga z+\de}\,,\qquad
C=\bpm\al & \be\\ \ga & \de\epm\,,\qquad
|C|=\al\de-\be\ga\neq 0\,,
\end{equation}
induces an action on $\p{n}$, mapping a polynomial $p(w)$ to
the polynomial $\pb(z)$ given by
\begin{equation}\label{eq.ron0}
\pb(z)=(\ga z+\de)^n\,p\Big(\frac{\al z+\be}{\ga z+\de}\Big).
\end{equation}
This defines an irreducible multiplier representation of $\GL(2)$ in
$\p{n}$,
\cite{Ol95}, which will be denoted by $\rho_{n,0}$. Since the
infinitesimal generators of this multiplier representation coincide with
the generators of $\g_n$, \cf\eqref{eq.gn}, it follows that the
representation $\rho_{n,0}$ induces an automorphism of the Lie algebra
$\hg_n$ spanned by the $\Je{n}$'s in \eqref{eq.J's}, \cite{GKO93bis}.
Performing the explicit scale transformation and change of variable,
\begin{equation}\label{eq.GL2onJ}
\Je n(w)\mapsto
(\ga z+\de)^n\,\Je n\Big(\frac{\al z+\be}{\ga z+\de}\Big)\,(\ga z+\de)^{-n}\,,
\end{equation}
we obtain:
$$
\bpm\Jp{n} \\ \J0{n} \\ \Jm\epm\mapsto\frac 1{|C|}
\bpm\al^2  &    2\al\be    & \be^2  \\
    \al\ga & \al\de+\be\ga & \be\de \\
    \ga^2  &    2\ga\de    & \de^2  \epm
\bpm\Jp{n} \\ \J0{n} \\ \Jm\epm\,.
$$
Therefore, the $\Je{n}$'s transform according to the representation
$\rho_{2,-1}=\rho_{2,0}\otimes\det^{-1}$ of $\GL(2)$, where $\det^{-1}$
is the reciprocal of the representation $\det:C\mapsto |C|$. It is
convenient at this stage to introduce a larger class of representations
of $\GL(2)$:

\begin{dfn}
Let $n\geq 0$, $i$ be integers. The (irreducible) multiplier
representation
$\rho_{n,i}$ of $\GL(2)$ on $\p n$ is defined by
$$
p(w)\mapsto\pb(z)=(\al\de-\be\ga)^i\,(\ga z+\de)^n\,
                  p\Big(\frac{\al z+\be}{\ga z+\de}\Big).
$$
\end{dfn}

We note the isomorphism between $\rho_{n,i}$ and
$\rho_{n,0}\otimes\det^i$. As shown in \cite{GKO93bis}, a second-degree
polynomial in the $\Je n$'s (in fact, any operator in $\pp n$ if $n\geq
2$, or any QES operator on the line modulo equivalence, according to
Theorems~\ref{thm.KaOl} and~\ref{thm.Tu}) may be written as
\begin{equation}\label{eq.p2J}
p_2(\Je n)=p\,\Dz^2+\big(q-\frac{n-1}2 p'\big)\,\Dz+
           r-\frac n2 q'+\frac{n(n-1)}{12}\,p''\,,
\end{equation}
where $p$, $q$, and $r$ are polynomials in $z$ of degrees $4$, $2$, and
$0$ respectively. The transformation of $p_2(\Je n)$ under the
action~\eqref{eq.GL2onJ} is easily described in terms of the triple
$(p,q,r)$, \cite{GKO93bis}:

\begin{lemma}\label{lem.pqr}
Let $p_2$ be a second-degree polynomial in the operators $\Je n(w)$,
determined by the triple $(p(w),q(w),r)$. Then, the transformed polynomial
$\pb_2$ under the $\GL(2)$ action~\eqref{eq.GL2onJ} is determined by the
triple $(\pb(z),\bar q(z),\bar r)$ given by
$$
\pb(z)=\frac{(\ga z+\de)^4}{|C|^2}\,p\Big(\frac{\al z+\be}{\ga
z+\de}\Big)\,,\qquad
\bar q(z)=\frac{(\ga z+\de)^2}{|C|}\,q\Big(\frac{\al z+\be}{\ga
z+\de}\Big)\,,\qquad
\bar r=r\,.
$$
\end{lemma}

Therefore, a second-degree polynomial $p_2$ in the $\Je n$'s transforms
according to the direct sum representation
$\rho_{4,-2}\oplus\rho_{2,-1}\oplus\rho_{0,0}$ under the $\GL(2)$
action~\eqref{eq.GL2onJ}. One can choose a particularly simple
representative of the $\GL(2)$ orbit generated by $p_2$ by placing the
polynomial $p$ (assumed to be real) in its associated triple $(p,q,r)$ in
canonical form,
\cite{Gu64}:

\begin{thm}\label{thm.canon}
Every non-zero quartic real polynomial $p(z)$ transforming under the
representation $\rho_{4,-2}$ of $\GL(2)$ is equivalent to one of the
following canonical forms:
\begin{xalignat}{2}
& 1)\;\nu(z^4+\tau z^2+1)\,,\;\tau\neq\pm 2,& & 5)\;\nu(z^2-1)\,,\notag\\
& 2)\;\nu(z^4+\tau z^2-1)\,,                & & 6)\;\nu z^2\,,\notag\\
& 3)\;\nu(z^2+1)^2\,,                       & & 7)\;z\,,\notag\\
& 4)\;\nu(z^2+1)\,,                         & & 8)\;1\,,
\end{xalignat}
where $\nu\neq 0$ and $\tau$ are real numbers.
\end{thm}

We now generalize these results to the matrix case. The induced action of
$\GL(2)$ on $\p{n-\De,n}$ analogous to \eqref{eq.ron0} is:
$$
\bpm p^1(w) \\ p^2(w) \epm\mapsto
\bpm \pb^1(z) \\ \pb^2(z) \epm=\hat U(z)
\bpm p^1(w(z)) \\ p^2(w(z)) \epm,
$$
where
$$
\hat U(z)=\diag\big((\ga z+\de)^{n-\De},(\ga z+\de)^n\big),
$$
and $w(z)$ is given by \eqref{eq.mobius}. This representation
of $\GL(2)$ in $\p{n-\De,n}$ is obviously isomorphic to
$\rho_{n-\De,0}\oplus\rho_{n,0}$. The following Lemma describes
the induced action on $\s_\De$:

\begin{lemma}\label{lem.roTJQ}
The action of $\GL(2)$ on $\s_\De$ given by
\begin{equation}\label{eq.GL2onX}
X(w)\mapsto\Xb(z)=\hat U(z)\,X\Big(\frac{\al z+\be}{\ga z+\de}\Big)\,\hat
U^{-1}(z)
\,,\qquad X\in\s_\De\,,
\end{equation}
defines a Lie superalgebra automorphism. The generators of $\s_\De$,
\cf\eqref{eq.TJQ}, transform according to the following irreducible
representations:
\begin{xalignat*}{2}
\{\Te\} &\rightarrow\rho_{2,-1}\,,     & \{J\}
&\rightarrow\rho_{0,0}\,,\\
\{\Qa\} &\rightarrow\rho_{\De,0}\,, & \{\Qab\}
&\rightarrow\rho_{\De,-\De}\,.
\end{xalignat*}
\end{lemma}

A straightforward generalization of equation~\eqref{eq.p2J} and of
Lemma~\ref{lem.pqr} to a second-degree polynomial in the $\Te$'s shows
that the (real) polynomial $p_4$ in \eqref{eq.T2} transforms according to
the representation $\rho_{4,-2}$ under the $\GL(2)$
action~\eqref{eq.GL2onX}. We will henceforth assume that $p_4$ is one of
the canonical forms given in Theorem~\ref{thm.canon}. The integral
\eqref{eq.xofz} and the inverse $z=\varphi^{-1}(x)$ can then be easily
computed for each of these canonical forms, \cite{GKO93}.

Before finishing this section, let us point out that
equations~\eqref{eq.Ux} and~\eqref{eq.hermitx} adopt a simpler form in
the variable $z$, because in that case only rational functions appear.
Explicitly, the equation for the gauge factor reads:
\begin{equation}\label{eq.Uz}
U'=U\Ah\,,\quad\text{with}\quad\Ah(z)=
\frac{A|_{\varphi(z)}}{\sqrt{p_4}}=\frac
1{2\,p_4}\big(A_1-\frac{1}{2}p_4'\big)\,,
\end{equation}
while $\Ut$ and $\Wt$ in equation~\eqref{eq.hermitx} are substituted by
$U$ and $W$, where
\begin{equation}\label{eq.W}
W=-A_0+p_4\Ah'+p_4\Ah^2+\frac 12\,p_4'\Ah\,.
\end{equation}
We now use Theorem~\ref{thm.TBK}, Lemma~\ref{lem.basis} and the
explicit form of the operators~\eqref{eq.TJQ} to compute
$\Ah$ and $A_0$ for the most general operator in $\pp{n-\De,n}$
of the form \eqref{eq.T2}, in the cases $\De=0,1$. We denote by
$p_n^\al$ the polynomial $\sum_{i=0}^n\al_iz^i$,
where $\al_i$ are arbitrary complex numbers. If the polynomial $p_4$ is
{\em not} one of the first three canonical forms, we obtain:
\begin{align}
\intertext{Case $\De=0$:}\label{eq.Delta0}
& \hspace{-4em} p_4\,\Ah = \bpm p_2^a & p_2^b \\ p_2^c & p_2^d \epm,\quad
  A_0 = \bpm \hat a_0-2n a_2 z & \hat b_0-2n b_2 z \\
             \hat c_0-2n c_2 z & \hat d_0-2n d_2 z \epm\!.\\
\intertext{Case $\De=1$:}\label{eq.Delta1} & \hspace{-4em} p_4\,\Ah =
\bpm p_2^a & p_1^b \\ p_3^c & p_2^d \epm,\quad
  A_0 = \bpm\hat a_0-2(n-1)a_2 z & -2n b_1 \\
            \hat c_0+\hat c_1 z-2(n-1)c_3 z^2 & \hat d_0-2n d_2 z\epm\!,
\end{align}
where $\hat a_0, \hat b_0, \hat c_0, \hat c_1$, and $\hat d_0$ are
arbitrary complex numbers. If $p_4$ is one of the first three canonical
forms, the following extra terms are present in $p_4\Ah$ and $A_0$:
\begin{align}
\hspace{-1em} p_4\,\Ah & \rightarrow -\diag\big((n-\De)\nu z^3,n\nu
z^3\big)\,,
\label{eq.extraA}\\
\hspace{-1em} A_0 & \rightarrow +\diag\big((n-\De)(n-\De-1)\nu
z^2,n(n-1)\nu z^2\big)\,.
\label{eq.extraA0}
\end{align}
\section{The Gauge Factor}\label{sec.gauge}

In this section we will deal with the differential
equation~\eqref{eq.Uz} for the gauge factor $U(z)$. As remarked in
Section~\ref{sec.sch}, this equation cannot be solved in closed
form for every $\Ah$. Equation~\eqref{eq.Uz} admits the formal
iterative solution
$$ U(z)=U_0\Big(1+\sum_{n=1}^\infty
\int^z
ds_1\,\int^{s_1}ds_2\,\dots\int^{s_{n-1}}ds_n\,\Ah(s_n)\dots
\Ah(s_2)\,\Ah(s_1)
\Big)\,,
$$
but this is of little practical use in checking the Hermiticity
of $V$. We will impose a certain condition on $\Ah$ so
that~\eqref{eq.Uz} can be explicitly solved, thus reducing the
number of parameters defining
$\Ah$, but still leaving us with room to construct relevant
examples of QES
\sch{} operators.

Denoting
$$ U=\bpm u_1 & u_2 \\ u_3 & u_4\epm\,,\qquad
\Ah=\bpm a & b \\ c & d\epm\,,
$$ we may write equation~\eqref{eq.Uz} as
\begin{equation}\label{eq.system}
\bpm u_1' \\ u_2'\epm=\bpm a & c \\ b & d\epm\bpm u_1 \\ u_2\epm\,,
\end{equation} with the same system for $(u_3,u_4)^t$.
Unfortunately, the associated scalar second order differential
equations for $u_1$ or $u_2$ are not any of the standard equations
of Mathematical Physics. We also note that the quotient $q=u_1/u_2$
satisfies the following Riccati equation:
\begin{equation}\label{eq.Riccati} q'=-bq^2+(a-d)q+c\,.
\end{equation} In fact, solving \eqref{eq.system} is equivalent to
solving \eqref{eq.Riccati}. Moreover, if we try to uncouple $\Ah$
by performing the linear transformation
$$
\bpm u_1 \\ u_2\epm=\bpm p & q \\ r & s\epm\bpm \bar u_1 \\ \bar
u_2\epm\,,
$$ we find that the transformed of $\Ah$ will be lower triangular
whenever
$q$ satisfies
\begin{equation}\label{eq.triangle} qs'-sq'+cs^2-bq^2+qs(a-d)=0\,,
\end{equation} with a similar condition to change $\Ah$ into upper
triangular form. If
$s$ is non-zero we can take $s=1$ without any loss of generality,
and equation~\eqref{eq.triangle} reduces to~\eqref{eq.Riccati}.
Unfortunately, none of these equations can be solved without
further assumptions.

However, if we restrict ourselves to matrices $\Ah$ satisfying the
equation
\begin{equation}\label{eq.comm0} [\Ah(z),\int^z_{z_0}
\Ah(s)\,ds]=0\,,
\end{equation} for some $z_0\in\Bbb R$, we can readily integrate
the gauge equation~\eqref{eq.Uz}. Recall that this condition on
$\Ah$ was indeed verified by the QES \sch{} operator found by
Shifman and Turbiner, \cite{ShTu89}. If
\eqref{eq.comm0} is satisfied, we shall say that we are in the {\em
commuting case}. In this case, the general solution of the gauge
equation is given by:
\begin{equation}\label{eq.sol}
U(z)=U_0\,\exp\int^z_{z_0}\Ah(s)\,ds\,,
\end{equation} where $U_0$ is in $\GL(2,\Bbb C)$. It is worth
mentioning in passing that we can use
\begin{equation}\label{eq.orbits}
\bpm 1 & 0 \\ \xi e^{i\phi} & \eta \epm\,,
\qquad \xi\geq 0\,,\quad \eta>0\,,\quad \phi\in[0,2\pi)
\end{equation} to parametrize the orbits of $\Bbb R^+\times\U(2)$
acting on $\GL(2,\Bbb C)$ by left multiplication. Since matrices in
the same orbit lead to equivalent \sch{} operators, we can take
$U_0$ in the form \eqref{eq.orbits}.

We shall be mainly concerned with the commuting case. We will make
use of the following elementary Lemma to describe the most general
form of
$\Ah$ in the commuting case:

\begin{lemma}\label{lem.comm} Let $M(z)$ be a $2\times 2$ matrix
satisfying the equation~\eqref{eq.comm0}. Then $M$ is of the form:
\begin{equation}\label{eq.M} M=f(z)\,M_0+g(z)\,,
\end{equation} where $f$ and $g$ are scalar functions, and $M_0$ is
a $2\times 2$ constant matrix.
\end{lemma}

With this Lemma in mind, looking at the expressions for $\Ah$
($\De=0,1$) that we obtained in Section~\ref{sec.canon},
\cf\eqref{eq.Delta0}--\eqref{eq.extraA}, we find that the most
general $\Ah$ satisfying \eqref{eq.comm0} is of the form
$$ p_4\,\Ah=\hat p_2(z)\,+\,\check A(z)\,,
$$ where, as usual, $\hat p_2$ denotes a second-degree polynomial
in $z$ with complex coefficients, and the matrix $\check A$ is
given in Table 1.

\hspace{-1.3em}
\begin{table}[h]
\begin{tabular}{r|cc} & $p_4$ in canonical forms 1--3 & $p_4$ in
canonical forms 4--8 \\[.3cm] \hline & & \\
$\De=0$\;&
$p_2\bpm \al & \be \\ \ga & \de \epm-n\nu z^3$ &
$\quad p_2\bpm \al & \be \\ \ga & \de \epm$ \\[.7cm]
$\De=1$\;&
$(p_2-\nu z^3) \bpm n-1 & 0 \\ \ga & n \epm$ & $\quad p_1 \bpm \al
& \be \\ \ga & \de \epm\!,\; p_2 \bpm \al & 0 \\ \ga & \de \epm\!,\;
p_3 \bpm 0 & 0 \\ \ga & 0 \epm$
\end{tabular}
\caption{Matrix $\check A(z)$ for the canonical forms 1--8 ($\al,
\be, \ga,\de\in\Bbb C$).}
\end{table}

Note that if $\De=1$ and $p_4$ is one of the first three canonical
forms, every Hamiltonian we can possibly obtain will be diagonal,
modulo equivalence.

Finally, let us remark that in the {\em non}-commuting case (that
is, when $[\Ah,\int^z \Ah]\neq0$), we may still be able to
integrate~\eqref{eq.Uz} explicitly by imposing other constraints on
$\Ah$, as \eg assuming it is uncoupled. Alternatively, if $p_4$ is
not any of the first three canonical forms, and we assume that the
columns (or rows) of $\Ah$ are proportional to each other (the
ratio of the respective entries being a constant), we can also
reduce~\eqref{eq.Uz} to quadratures. Unfortunately, we have not
been able to find any interesting examples of QES Hamiltonians in
the non-commuting case.

\section{Examples}\label{sec.examples}

In this final section we exhibit some new examples of spin $1/2$
\sch{} operators equivalent to a PVSP operator of the
form~\eqref{eq.T2}. In the previous section we have seen how, by
restricting ourselves to the commuting case, we were able to
integrate equation~\eqref{eq.Uz} explicitly. This is not, however,
the end of the problem, for we must still check that the matrix
\begin{equation}\label{eq.pot} V=UWU^{-1}\big|_{\varphi^{-1}(x)}\,,
\end{equation} with $W$ given by \eqref{eq.W}, is self-adjoint.
Again, if we start with the most general PVSP operator of the
form~\eqref{eq.T2}, the algebraic constraints imposed by the
condition $V=V^\dagger$ are too complicated to be solved in full
generality, even if we limit ourselves to the commuting case. This
situation is completely analogous to what we find when trying to
construct scalar QES
\sch{} operators in more than one spatial dimension, \cite{Sh89},
\cite{GKO93}. We have thus no choice but to look for particular
examples. We now present some relevant examples of QES spin $1/2$
Hamiltonians for the commuting case. In the first two examples we
take $\De=0$, while in the remaining ones we assume that $\De=1$.
Finally, let us point out that many otherwise interesting examples
are often reduced to trivial ones after we impose the square
integrability condition on the eigenfunctions.\\

\noindent{\bf Example 1}. Consider the six-parameter PVSP operator
$T$ given by
\begin{align*} -T=\: & \Tm\T0+(a_2+d_2)\,\Tp+(a_2-d_2)\,\tilde
J\Tp+2a_1\T0+
\bigl(2a_0+\frac 12(n-1)\bigr)\,\Tm
\\    & +(Q_0+\Qb_0)(2\,b_2\,\Tp+\hat b_0)\,,
\end{align*} where all the parameters are real numbers coinciding
with those appearing in equation~\eqref{eq.Delta0}, and $b_2\neq
0$. Note that we are in case 7 of Theorem~\ref{thm.canon}
($p_4=z$), and so $z=x^2/4$. The gauge factor
$U(z)$ is given by
$$ U(z)=f(z) \bpm \La\,\cosh u+(a_2-d_2)\,\sinh u & 2\,b_2\,\sinh u
\\
               2\,b_2\,\sinh u & \La\,\cosh u+(d_2-a_2)\,\sinh u
\epm\,,
$$ with
$$ f(z)=z^{a_0}\,\exp\big(a_1 z+\frac 14(a_2+d_2)z^2\big)\,,
\qquad u=\frac\La4 z^2\,,
\qquad \La^2=(a_2-d_2)^2+4\,b_2^2\,,
$$ where we have taken $U_0=\La$ in equation~\eqref{eq.sol}.
Using~\eqref{eq.pot}, we obtain a potential $V(x)$ with entries
given by (see~\eqref{eq.V}):
\begin{align*} & v_1=\al_{-2}x^{-2}+\al_0+\al_2 x^2+\al_4 x^4+\al_6
x^6\,,\\ & v_2=\de_{-2}x^{-2}+\de_0+\de_2 x^2+\de_4 x^4+\de_6
x^6\,,\\ & v=\be_0+\be_2 x^2+\be_4 x^4+\be_6 x^6\,,
\end{align*} where
\begin{align*} & \al_{-2}=2a_0(2a_0-1)\,,\qquad
\al_0=-\frac{a_2\hat b_0}{b_2}\,,\qquad
  \al_2=\frac{a_1^2}4+\frac{a_2}2\big(a_0+n+\frac 34\big)\,,\\ &
\al_4=\frac{a_1a_2}8\,,\qquad \al_6=\frac{a_2^2+b_2^2}{64}\,,\\ &
\be_0=-\hat b_0\,,\qquad \be_2=\frac {b_2}2\big(a_0+n+\frac
34\big)\,,\qquad
  \be_4=\frac{a_1b_2}8\,,\qquad \be_6=\frac{b_2(a_2+d_2)}{64}\,,
\end{align*} while each $\de_n$ is given by the corresponding
$\al_n$ replacing $a_2$ by $d_2$. We have ignored a constant
multiple of the identity in $V$, which is equivalent to fixing a
new origin in the energy scale (this will be done for the subsequent
examples without further notice).

We note that the potential $V$ is singular at $x=0$ unless $a_0$ is
either $0$ or $1/2$. Let us introduce the parameter $\la=2a_0-1$, in
terms of which we have $\al_{-2}=\la(\la+1)$. If $\la$ is a
non-negative integer $l$, we may regard
\begin{equation}\label{eq.sch1} (-\Dx^2+V(x)-E\,)\PSI(x)=0\,,\qquad
0<x<\infty\,,
\end{equation} as the radial equation obtained after separating
variables in the three-di\-men\-sion\-al \sch{} equation with a
spherically symmetric Hamiltonian given by
$$
\hat H=-\De+U(r)\,,\qquad\text{with}\qquad
U(x)=V(x)-\frac{l(l+1)}{x^2}\,,
$$ where $\De$ denotes the usual flat Laplacian. Given a
non-negative integer $l$ and a spherical harmonic
$Y_{lm}(\theta,\phi)$, if $\PSI$ is an eigenfunction for the
equation~\eqref{eq.sch1} satisfying
\begin{equation}\label{eq.bc1}
\lim_{x\rightarrow 0^+}\PSI(x)=0\,,
\end{equation} then
$$
\hat\Psi(r,\theta,\phi)=\frac{\PSI(r)}r\,Y_{lm}(\theta,\phi)
$$ will be an eigenfunction for $\hat H$ with angular momentum $l$.
If $\la$ is not a non-negative integer, we shall
consider~\eqref{eq.sch1} as the radial equation for the singular
potential $U(r)=V(r)$ at zero angular momentum. The potential
$U(r)$ is physically meaningful, in the sense that the Hamiltonian
$\hat H$ admits self-adjoint extensions and its spectrum is bounded
from below, whenever $\la\neq -1/2$, \cite{GaPa90},
\cite{GKO93bis}. The boundary condition~\eqref{eq.bc1} must be
satisfied in the singular case for {\em all} values of $\la$. This
boundary condition is verified if and only if
$a_0>0$. Finally, the additional conditions
$$ a_2\,d_2>b_2^2\,,\qquad a_2<0\,,
$$ ensure that $\PSI$ lies in $L^2(I)\oplus L^2(I)$, where
$I=[0,\infty)$ in the singular case or at zero angular momentum, or
$I=\Bbb R$ in the non-singular one-dimensional case.\\

\noindent{\bf Example 2}. As our second example, we take
\begin{align*} -T=\: & (\Tm)^2+(a_1+d_1)\,\T0+(a_1-d_1)\,\tilde
J\T0+2a_0\,\Tm\\
      & +(b_1 Q_0+b_1^\ast \Qb_0)(2\,\T0+n-\mu)+\frac
12(n-\mu)(a_1-d_1)\tilde J\,,
\end{align*} where the coefficients are real, excepting $b_1\neq 0$
which is complex. In this example $p_4=1$ and thus $z=x$ (case 8 of
Theorem~\ref{thm.canon}). We choose the following gauge factor:
$$ U(z)=f(z) \bpm \La\,\cosh u+(a_1-d_1)\,\sinh u &
2\,b_1^\ast\,\sinh u \\
               2\,b_1\,\sinh u & \La\,\cosh u+(d_1-a_1)\,\sinh u
\epm\,,
$$ with
$$ f(z)=\exp\big(a_0 z+\frac 14(a_1+d_1)z^2\big)\,,
\qquad u=\frac\La4 z^2\,,
\qquad \La^2=(a_1-d_1)^2+4\,|b_1|^2\,.
$$ We obtain the potential with entries given by
\begin{align*} & v_1=a_1(\mu+1)+2 a_0
a_1\,x+(a_1^2+|b_1|^2)\,x^2\,,\\ & v_2=d_1(\mu+1)+2 a_0
d_1\,x+(d_1^2+|b_1|^2)\,x^2\,,\\ & v=b_1\big(\mu+1+2
a_0\,x+(a_1+d_1)\,x^2\big)\,.
\end{align*} The eigenfunctions $\PSI(x)$ are square-integrable
provided
$$ a_1\,d_1>|b_1|^2\,,\qquad a_1<0\,.
$$ Since the parameter $n$ does not appear in the gauge factor or
in the potential, the Hamiltonian preserves $U\cdot\p{n,n}$ for
{\em arbitrary} $n$. (In the literature, an operator with this
property is usually referred to as exactly solvable, \cite{Tu92},
\cite{Tu92bis}, \cite{BrKo93}.) Note that if $\mu=2 n$ we could
also obtain $T$ from the PVSP operators~\eqref{eq.TJQ} with $\De=1$,
but in this case the Hamiltonian would preserve $U\cdot\p{m-1,m}$
only for $m=n$.\\

\noindent{\bf Example 3}. From now on we take $\De=1$. Let $T$ be
the four-parameter PVSP operator given by
\begin{align*} -T=\: & (\T0)^2+2 a_2\Tp+2(n+1)\T0-2\,J\T0+2
b_1\,\Qb_0+2 b_0\,\Qb_1-2 b_1\,Q_0\T0\\ & -2 b_0\,Q_0\Tm-\bigl(4
a_2 b_0+(3n+1)b_1\bigr)\,Q_0-4 a_2 b_1 Q_1-(2\hat d_0+n+\frac
12)J\,,
\end{align*} with all the parameters real. Since $p_4=z^2$ we are
in case 6 of Theorem~\ref{thm.canon}. Solving
equation~\eqref{eq.xofz} for $z$, we obtain
$z=e^x$. The gauge factor reads:
$$ U(z)=\sqrt z\,e^{a_2 z}\bpm \cos u & \sin u \\ -\sin u & \cos u
\epm\,,
\qquad\text{where}\qquad u=-\frac{b_0}z+b_1\log z\,.
$$ The potential is given by
\begin{align*} & v_j=-b_0^2\,e^{-2x}-2 b_0
b_1\,e^{-x}+(2n+1)a_2\,e^x+a_2^2\,e^{2x}\\ & \qquad
+(-1)^j\big(\al(x)\cos 2\tilde u\,-\,\be(x)\sin 2\tilde u\big)\,,
\qquad\qquad\qquad j=1,2\\[.2cm] & v=\al(x)\sin 2\tilde
u\,+\,\be(x)\cos 2\tilde u\,,\vspace{.2cm}
\end{align*} where $\;\tilde u=b_1\,x-b_0\,e^{-x}$, and
$$
\al(x)=-\frac{\hat d_0}2+a_2\,e^x\,,\qquad
\be(x)=(2n+1)\,b_1+2a_2(b_0+b_1\,e^x)\,.
$$ It may be easily verified that the expected value of the
potential is bounded from below, \ie
\begin{equation}\label{eq.bound}
\langle\PSI,V\PSI\rangle\geq c\,\|\PSI\|^2\,,
\qquad\text{with}\quad\PSI\in L^2(\Bbb R)\oplus L^2(\Bbb R)\,,
\end{equation} for some $c\in\Bbb R$, if and only if $b_0=0$.
(Note, however, that even in this case the amplitude of the
oscillations of $v(x)$ tends to infinity as
$x\rightarrow+\infty$.) Finally, the condition $a_2<0$ is necessary
and sufficient to ensure the square integrability of the
eigenfunctions $\PSI(x)$.\\

\noindent{\bf Example 4}. As our last example, we consider:
\begin{align*} -T=\: & \Tm\T0+2 a_1\T0+(2 a_0+n-\frac
12)\,\Tm-J\Tm+2 b_1\,\Qb_0-2 b_1\,Q_0\T0\\ & -b_1\,(4
a_0+3n+1)\,Q_0-4 a_1 b_1 Q_1+2(2\hat a_0-a_1)J\,,
\end{align*} where all the coefficients are real, and $b_1\neq 0$.
Since $p_4=z$ (case 7), we have $z=x^2/4$. The gauge factor is
chosen as follows:
$$ U(z)=z^{a_0}\,e^{a_1 z}
\bpm \cos b_1 z & \sin b_1 z \\ -\sin b_1 z & \cos b_1 z \epm\,.
$$ The entries of the potential $V(x)$ are given by
\begin{align*} & v_j=\frac{2a_0(2a_0-1)}{x^2}+\frac
14\,(a_1^2-b_1^2)\,x^2
      +(-1)^j\big(\hat a_0\,\cos\frac{b_1
x^2}2\,-\,\al(x)\,\sin\frac{b_1 x^2}2\big)
\,,\\[.2cm] & v=\hat a_0\,\sin\frac{b_1
x^2}2\,+\,\al(x)\,\cos\frac{b_1 x^2}2\,,\vspace{.2cm}
\end{align*} with $j=1,2$, and $\al(x)$ is defined by
$$
\al(x)=\frac{b_1}2\,(4a_0+4n+1+a_1 x^2)\,.
$$ We first note that the potential is singular at the origin
unless $a_0=0,\,1/2$. This situation is completely analogous to
that of Example 1, so it will not be further discussed. The
expected value of the potential is bounded from below, that is,
equation~\eqref{eq.bound} holds, if and only if
$$
\left|\frac{a_1}{b_1}\right|>1+\sqrt 2\,.
$$ Finally, the conditions
$$ a_0\geq 0\,,\qquad a_1<0\,,
$$ guarantee the square integrability of the eigenfunctions
$\PSI(x)$.\\

\end{document}